# Adsorbates as a charge-carrier reservoir for electrostatic carrier doping to graphene

Ryo Nouchi[1,2]* and Kei-ichiro Ikeda[1]

[1]*Department of Physics and Electronics, Osaka Prefecture University, Sakai 599-8570, Japan*

[2]*PRESTO, Japan Science and Technology Agency, Kawaguchi 332-0012, Japan*

E-mail: r-nouchi@pe.osakafu-u.ac.jp

A charge-carrier reservoir is necessary for electrostatic control of the carrier concentration in a solid. The source/drain electrodes serve as carrier reservoirs in a field-effect transistor, but it is still unknown what serves as a reservoir in a technique based on a polar self-assembled monolayer formed underneath a solid to be controlled. Here, the carrier-doping level of isolated single-layer graphene was found to be the same as that of the single-layer part in a flake containing multilayer graphene, indicating that the multilayer part is not a dominant carrier reservoir but adsorbates like oxygen and water serve as a dominant reservoir.

The ultrathin body of two-dimensional (2D) materials obtained by the exfoliation of layered compounds allows control of their carrier concentrations throughout the entire body. The first example on an archetypal 2D material, graphene, has been an electrostatic doping technique using a field-effect-transistor (FET) configuration,[1] which exploits the capacitive coupling of two layers, a graphene flake and metal (the gate electrode), separated by a dielectric. Since this report on electrostatic doping, charge-transfer doping exploiting an interface with another substance has been intensively studied with a variety of charge-donating substances, including alkaline metals,[2,3] metal electrodes,[4-7] small molecules,[8-11] and inorganic insulators,[12] etc.

More recently, another interface-type doping technique that is independent from the direct charge transfer has been introduced to control the carrier concentration of graphene.[13] This technique employs chemical modification of the supporting substrate surface with polar molecules. The resultant self-assembled monolayer (SAM) formed on the supporting substrate possesses an electric double layer that originates from the electric dipoles of the constituent molecules of the SAM. Even without the direct charge transfer, the SAM with an electric double layer can act the same as a polarized dielectric in the FET-type electrostatic doping. This technique has been employed to control the threshold voltage of organic FETs,[14,15] and more recently to control the carrier concentration of 2D materials, such as graphene[13] and transition metal dichalcogenides.[16] Therefore, the polar-SAM-based doping technique is quite similar to the FET-based electrostatic doping in terms of the operation principle, though it displays structural similarity to charge-transfer doping, as both techniques exploit an interface with another substance.

For the proper operation of the electrostatic doping technique (the FET- and SAM-based techniques), a charge-carrier reservoir is necessary to supply the carriers to the 2D material layer. As depicted in Fig. 1(a), the electrode metals (source/drain electrodes) act as the reservoir in the FET-based technique. However, the report on the SAM-based technique for mechanically exfoliated graphene[13] showed that the carrier concentration control can be achieved without any electrodes on the graphene layer, and proposed that the multilayer part in the same flake can act as the reservoir. In the case of mechanically exfoliated graphene, single layer graphene (SLG) is frequently found inside a thickness-varying flake. Thus, in the SAM-based technique, the multilayer graphene (MLG) part in the same flake is expected to act as a charge-carrier reservoir even without electrode metals, as shown in Fig. 1(b). This concept of the intra-flake reservoir is not restricted to 2D semimetals (i.e., graphene and graphite) and should work also with 2D semiconductors because charge-donating impurities (donors or acceptors), whether intentionally and unintentionally, exist in the 2D semiconductors. This concept is the vital point for understanding the mechanism of the SAM-based doping technique. However, the direct experimental verification of the concept has not yet been provided.

In this paper, experimental proof that the MLG part is not a dominant carrier reservoir is provided via detailed analyses using Raman scattering spectroscopy. The Fermi energies of graphene flakes mechanically exfoliated on SAM-modified $SiO_2$ substrates are extracted from the acquired Raman spectra, and are found to be similar values between the isolated SLG flake and thickness-varying flake. This strongly suggests that the charge-carrier doping is accomplished by carrier injection from a substance other than the intra-flake MLG part. Adsorbates (on/underneath SLG) instead of the MLG part are proposed to act as a dominant charge-carrier reservoir.

A Si wafer with a 300-nm-thick thermal oxide layer on top was used as the supporting substrate. Immediately after the $SiO_2$ surface was cleaned by an oxygen plasma treatment (Harrick Plasma, PDC-32G), the substrate was immersed in a 2 wt% hexane solution of *n*-propyltriethoxysilane (Tokyo Chemical Industry, purity > 98%) or 1H,1H,2H,2H-perfluorooctyltriethoxysilane (Tokyo Chemical Industry, purity > 95%) for 20 h in ambient conditions. The substrate was then cleaned with pure hexane in an ultrasonic bath. The resultant SAMs of these two molecules are hereafter called $CH_3$-SAM and F-SAM, respectively. Immediately after the SAM formation, graphene flakes were fabricated onto the SAM-modified substrate by mechanical exfoliation using adhesive tape.[17] Figure 1(c) shows the direction of the permanent electric dipole of the molecules and a schematic illustration of the fabricated samples. The characterization of the flakes was performed in ambient air using a Raman microscope (Nanophoton, Raman-DM) equipped with a 532-nm laser. Acquired Raman scattering spectra were calibrated using a Si-related peak at 520 $cm^{-1}$.

Figure 2 shows the Raman scattering spectra of SLG formed on $CH_3$- and F-SAM. Two types of flakes are shown in Fig. 2, namely, an isolated SLG flake and a thickness-varying flake. The former sample does not contain an MLG part within the flake, and thus an MLG part is excluded as a charge-carrier reservoir. The D band (~1340 $cm^{-1}$), which is indicative of the presence of defects, was found to be very small in these flakes. Although the isolated SLG flake on $CH_3$-SAM shows a recognizable D peak, the ratio of the D band to the G band (~1590 $cm^{-1}$) was calculated to be only 0.02. These facts indicate that SAM-induced damages to the graphene layers upon were negligible in these flakes. The G band of graphene on F-SAM shows a larger Raman shift than that on $CH_3$-SAM, and the ratio of the 2D band (~2680 $cm^{-1}$) to the G band is smaller in graphene on F-SAM than that on $CH_3$-SAM; both of these facts are indicative of a higher level of carrier doping in graphene on F-SAM.[18] The doped carrier type is expected to be dependent on the polarity of the adjacent pole of the electric dipoles. For example, in the case of F-SAM, the polarity of the pole adjacent to the graphene flake is negative because of the higher electron affinity of fluorine atoms, which induces doping of the charge carriers with a counter charge, i.e., holes to graphene [Fig. 1(b)]. This situation corresponds to the negative gate-voltage application in the FET-based electrostatic doping [Fig. 1(a)]. The doping ability of the SAM-based technique has been confirmed by combining with the FET-based technique as a shift of the transfer characteristics of graphene FETs.[19,20] Surprisingly, the trend is independent on whether the analyzed flake contains an MLG part or not. This indicates that an MLG part within the graphene flake is not a dominant reservoir of charge carriers.

The doping level can be determined by inspecting Raman scattering spectra. The ground for the determination is how the position and shape of characteristic Raman peaks change in response to the gate voltage application with an FET configuration, which has been given by several groups.[18,21] In these studies, systematic changes in the Raman features were observed with a clear correlation to transfer characteristics of graphene FETs. In addition, the changes in the Raman features were in very good agreement with theoretical predictions. Although the doping level controlled by the SAM-based technique cannot be confirmed by FET characteristics because of the absence of metallic electrodes, Raman scattering spectroscopy instead allows us to accurately determine the doping level of each SLG flake.

The specific doping level at each point in the flakes can be determined by analyzing the 2D–G correlation.[21,22] The analysis relies on the different trajectories of strain- and doping-induced shifts in an $\omega_{2D}$–$\omega_G$ plot, where $\omega_{2D}$ and $\omega_G$ are the peak wavenumbers of the 2D and G bands, respectively. As depicted in Fig. 3(a), slopes of the trajectories have been reported to be 2.2 for a uniaxial strain[22] and 0.55 (0.2) for hole (electron) doping.[21] The origin of the linear trajectories has been reported to be $(\omega_{2D}^0, \omega_G^0) = (2676.9, 1581.6)$ $[cm^{-1}]$ for an excitation laser wavelength of 514 nm.[22] Considering a dispersive nature of the 2D band (the shift of $\omega_{2D}^0$ by a change in an excitation energy is 88 $cm^{-1}/eV$),[23] the origin for the present study with an excitation at 532 nm becomes $(\omega_G^0, \omega_{2D}^0) = (1581.6, 2669.7)$ $[cm^{-1}]$. Figure 3(a) shows the 2D–G correlation of graphene formed on $CH_3$- and F-SAM modified substrates. The analyzed area is the same as the red square region shown in the optical micrographs in Fig. 2. In Fig. 3(a), data points from the SLG on $CH_3$-SAM are located very close to the strain line, indicating the almost undoped nature of the SLG on $CH_3$-SAM.

In contrast, data points from the SLG formed on F-SAM are distant from the strain line. These facts also indicate the higher doping level of SLG on F-SAM.

As schematically shown in Fig. 3(b), a vector-decomposition analysis using the strain and doping lines is necessary to determine, for example, the $\omega_G$ value corresponding to the no-strain condition (i.e., the $\omega_G$ value determined solely by the hole doping level). The no-strain values are necessary to quantify the doping level, i.e., to determine the Fermi level. However, there are two possible doping lines for the vector-decomposition analysis. Thus, the doped carrier type (electron or hole) should be determined prior to performing the analysis. Many Raman features, such as $\omega_G$, the full width at half maximum (FWHM) of the G band, and the intensity ratio of the 2D peak to the G peak, are nearly symmetrical with respect to electron/hole doping.[18,21] $\omega_{2D}$ is known to possess electron–hole asymmetry in the high-doping regime;[18,21] however, it shows an almost symmetrical change within $|E_F| \lesssim 200$ meV, where $E_F$ is the Fermi level relative to the Dirac point,[21] and cannot be used for the carrier-type determination in the low-doping regime. The FWHM of the 2D band shows a very weak monotonic increase as $E_F$ increases from the hole- to the electron-doping regime. Figure 3(c) shows the plots between the FWHM of the 2D band and as-measured $\omega_G$ values. The SLG parts on F-SAM shows a decrease trend as $\omega_G$ increases, indicating hole doping as expected from the direction of the electric dipole [Fig. 1(c)]. In contrast, those on CH3-SAM show almost no dependency on $\omega_G$, indicating the coexistence of the electron- and hole-doped regions; this result is consistent with the almost undoped nature, as demonstrated in Fig. 3(a).

The determined carrier type (hole for SLG on F-SAM) was used to perform the vector-decomposition analysis. The extracted $\omega_G$ values corresponding to the no-strain condition are collected in Fig. 3(d) as a histogram. In the histogram, the $\omega_G$ values for SLG on CH3-SAM were also extracted under assumption of hole doping; however, the nearly undoped nature guarantees a small error in the extracted values with regard to the assumption of the doped carrier type. The corrected $\omega_G$ values were determined to be $1581.3 \pm 0.5$ cm$^{-1}$ (isolated SLG on CH3-SAM), $1581.8 \pm 0.9$ cm$^{-1}$ (SLG part in an MLG-containing flake on CH3-SAM), $1588.0 \pm 0.2$ cm$^{-1}$ (isolated SLG on F-SAM), and $1588.1 \pm 0.2$ cm$^{-1}$ (SLG part in an MLG-containing flake on F-SAM). The extracted $\omega_G$ value with no strain was found to be almost the same for the isolated SLG and the SLG part in a flake containing an MLG part. The corrected $\omega_G$ values are necessary to quantitatively determine the doping level. $E_F$ relative to the Dirac point can be determined by using the corrected $\omega_G$ in cm$^{-1}$, as $E_F = -18(\omega_G - \omega_G^0) - 83$ [meV] for hole doping within 100 meV $\lesssim |E_F| \lesssim 500$–600 meV.[21] The $\omega_G$ values within $|E_F| \lesssim 100$ meV are known to be nearly constant[21] because of spatial $E_F$

fluctuations caused by electron–hole puddles.[24,25] The $E_F$ values of the SLG on F-SAM were extracted to be $-198 \pm 3$ meV and $-201 \pm 3$ meV for the isolated SLG and the SLG part within a flake containing MLG, respectively. All of these considerations show that the doping level does not depend on whether the analyzed SLG part is isolated or attached to an MLG part.

The discussions made above indicate that the intra-flake MLG part is not a dominant carrier reservoir for SAM-based electrostatic doping. Conceivable reservoirs other than the MLG part are the substrate, SAM molecule, and adsorbed foreign molecules. The substrate, $SiO_2$, is a wide-gap insulator, and direct charge transfer to SLG is energetically unfavorable; thus, the substrate can be excluded as the reservoir in our sample structure. The SAM molecules used in this study have been calculated to induce no direct charge transfer to graphene,[13] and can thus also be excluded. Therefore, we propose that the main reservoir of charge carriers comprises foreign molecules adsorbed on/underneath graphene, as schematically shown in Fig. 4. The foreign molecules should include oxygen, water, and resist residues. Among them, the former two molecules are known to dope holes via the oxygen/water redox couple.[26] In the present study, Raman scattering spectra were acquired in ambient air, from which oxygen and water were provided. The molecules involved in the redox couple, which are omnipresent in ambient air, are responsible for the hole doping to SLG on F-SAM. To clarify what kind of adsorbates act as the reservoir, further experiments in a controlled gas environment should be conducted: e.g., preparation of graphene flakes in an inert atmosphere, Raman spectroscopy measurements performed in an inert atmosphere after annealing in the same atmosphere, etc.

In conclusion, a charge-carrier reservoir in an electrostatic carrier-doping technique was investigated using graphene flakes formed on polar-SAM-modified substrates. In this technique, which we call molecular gating, charge carriers possessing the countercharge to the adjacent pole of the SAM dipole are electrostatically doped to SLG, which was evidenced by a shift in the G peak of the Raman scattering spectra. Whereas the charge-carrier reservoir in FET-based doping is electrode metals (source/drain contacts), that in SAM-based doping has been believed to be an MLG part so far. We found that the doping level of an isolated SLG flake was nearly the same as that of an SLG part in a flake containing the MLG part. This result indicates that the MLG part is not a dominant carrier reservoir in the present case. We propose that adsorbates, such as oxygen and water molecules, can instead act as a charge-carrier reservoir. For its simplicity in a sample structure (no need for wiring and applying gate voltages), the molecular gating technique provides an easy way to control the charge-carrier concentrations of graphene. In principle, this technique is applicable to all 2D materials, which clearly indicates the significance of the technique. The present study is an important step toward understanding the background mechanism of the molecular gating technique.

**Acknowledgments**

This work was supported by JSPS KAKENHI Grant Numbers JP26107531, JP16H00921, JP17H01040, and JP19H02561; and JST, PRESTO Grant Number JP17939060.

## Figure Captions

**Fig. 1.** Electrostatic control of charge-carrier concentration in graphene. The charge-carrier reservoir comprises metal electrodes in the FET-based technique (a), and was proposed to be an MLG part in the SAM-based technique (b). (c) Schematic diagram of the SAM-based technique. The doped carrier type is determined by the orientation of the permanent electric dipole of the SAM molecules.

**Fig. 2.** Raman scattering spectra of an SLG part in flakes formed on $CH_3$-SAM and F-SAM. The spectra were averaged in the red square regions in the optical micrographs, where the dashed white lines indicate the SLG part. The spectra are normalized to the peak intensity of the G peaks. The insets show enlarged views near the G and 2D band regions.

**Fig. 3.** 2D–G correlation analysis for determination of the doping level. The analyzed areas are the same as the red squares shown in Fig. 2. (a) 2D–G correlation plot of as-measured wavenumbers of the G and 2D peaks. (b) Schematic diagram of the vector-decomposition analysis. (c) FWHM of the 2D peak as a function of as-measured $\omega_G$. Negative slopes suggests that the SLG parts are hole-doped. (d) Histogram of G peak wavenumbers corresponding to no-strain condition. The extracted values were determined using the hole doping line.

**Fig. 4.** Adsorbates proposed as a charge-carrier reservoir for the SAM-based electrostatic control of carrier concentration in graphene.

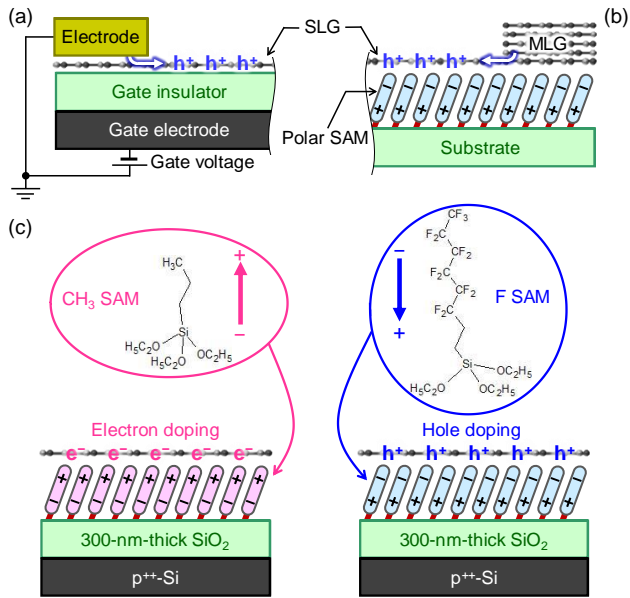

Fig.1.

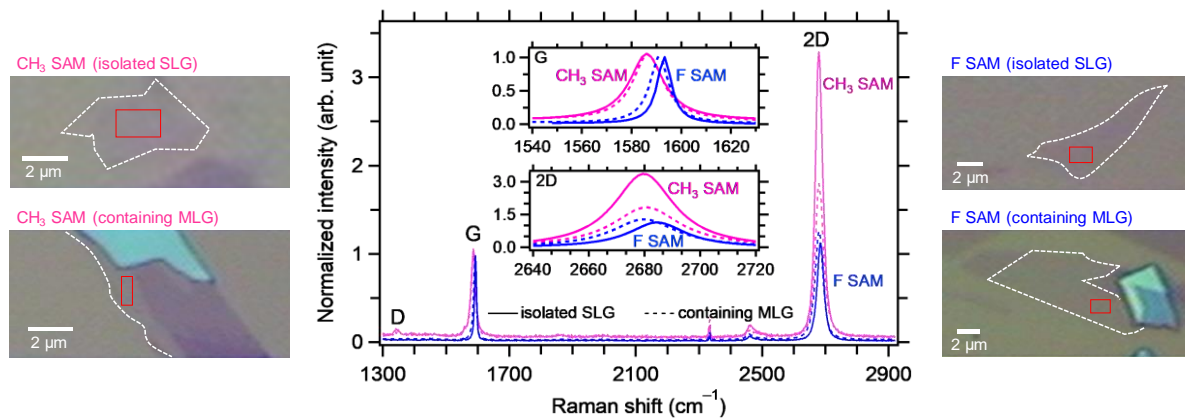

Fig. 2.

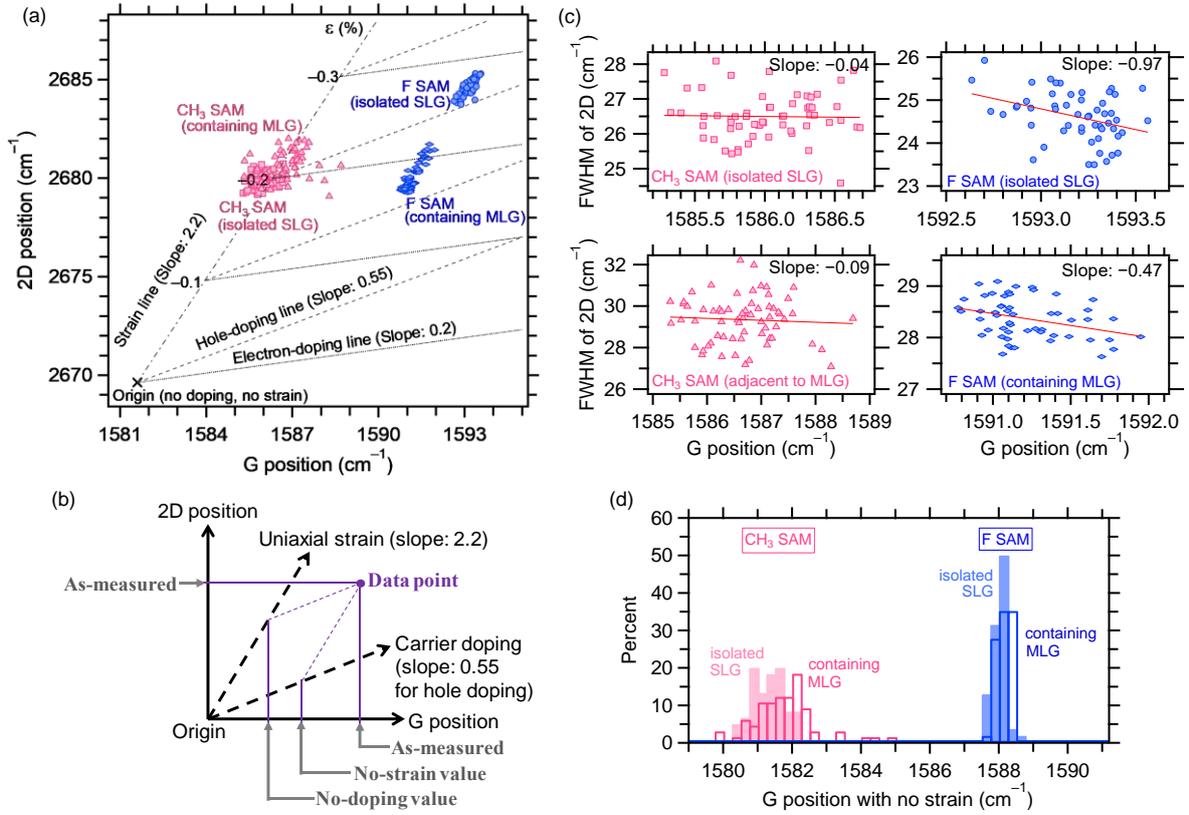

Fig. 3.

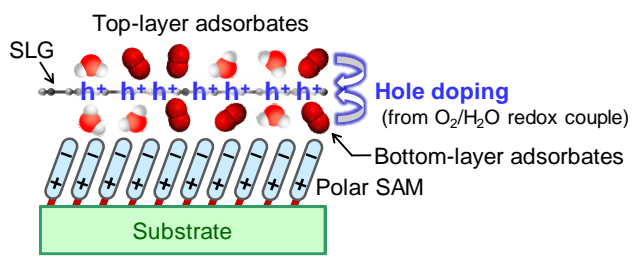

Fig. 4.